\documentclass[a4paper,fleqn,usenatbib,useAMS,usedcolumn]{mnras}

\usepackage[T1]{fontenc}
\usepackage{ae,aecompl}

\usepackage{graphicx}	
\usepackage[fleqn]{amsmath}	
\usepackage{amssymb}	
\usepackage{gensymb}
\usepackage[nolist]{acronym}
\usepackage{ulem}

\usepackage{etoolbox}

\acrodef{GW}{gravitational wave}
\acrodef{BH}{black hole}
\acrodef{BBH}{binary black hole}
\acrodef{NSBH}{neutron star--black hole}
\acrodef{aLIGO}{Advanced LIGO}
\acrodef{AdV}{Advanced Virgo}
\acrodef{PN}{post-Newtonian}
\acrodef{BNS}{binary neutron star}
\acrodef{PE}{parameter estimation}
\acrodef{SNR}{signal-to-noise ratio}
\acrodef{IMF}{initial-mass function}
\acrodef{CE}{common-envelope}
\acrodef{MCMC}{Markov-chain Monte Carlo}
\acrodef{ZAMS}{zero-age main sequence}

\newcommand{\sub}[1]{\ensuremath{_\mathrm{#1}}}

\newcommand{\vect}[1]{\boldsymbol{#1}}
\newcommand{\uvect}[1]{\hat{\boldsymbol{#1}}}

\newcommand{\change}[1]{{\color{black}#1}}

\title[BBH spin misalignments]{Hierarchical analysis of gravitational-wave measurements of binary black hole spin--orbit misalignments}

\author[S. Stevenson et al.]{\parbox{\textwidth}{
Simon Stevenson,$^{1}$\thanks{E-mail: simon.stevenson@ligo.org }
Christopher P.\ L.\ Berry,$^{1}$
and Ilya Mandel$^{1}$}
\vspace{0.2cm}\\
\parbox{\textwidth}{$^{1}$School of Physics and Astronomy and Institute of Gravitational Wave Astronomy, University of Birmingham, Edgbaston, Birmingham B15 2TT, United Kingdom}\vspace{0.2cm}
}

\date{\today}

\pubyear{2017}

\begin{document}
\label{firstpage}
\pagerange{\pageref{firstpage}--\pageref{lastpage}}
\maketitle

\begin{abstract}

Binary black holes may form both through isolated binary evolution and through dynamical interactions in dense stellar environments. \change{The formation channel leaves an imprint on the alignment between the black hole spins and the orbital angular momentum.} Gravitational waves from these systems directly encode information about the spin--orbit misalignment angles, allowing them to be (weakly) constrained. Identifying sub-populations of spinning binary black holes will inform us about compact binary formation and evolution. 
We simulate a mixed population of binary black holes with spin--orbit misalignments modelled under a range of assumptions. We then develop a hierarchical analysis and apply it to mock gravitational-wave observations of these populations. \change{Assuming a population with dimensionless spin magnitudes of $\chi = 0.7$,} we show that tens of observations will make it possible to distinguish the presence of subpopulations of coalescing binary black holes based on their spin orientations.  With $100$ observations it will be possible to infer the relative fraction of coalescing binary black holes with isotropic spin directions (corresponding to dynamical formation \change{in our models}) with a fractional uncertainty of $\sim 40\%$.  Meanwhile, only $\sim 5$ observations are sufficient to distinguish between extreme models---all binary black holes either having exactly aligned spins or isotropic spin directions.  

\end{abstract}

\begin{keywords}
black hole physics  --  gravitational waves  --  methods: data analysis  --  stars: evolution  -- stars: black hole
\end{keywords}



\section{Introduction}
\label{sec:Introduction}

Compact binaries containing two stellar-mass \acp{BH} can form as the end point of isolated binary evolution, or via dynamical interactions in dense stellar environments \citep[see, e.g.,][for a review]{Mandel:2009nx,TheLIGOScientific:2016htt}. These \acp{BBH} are a promising source of \acp{GW} for ground-based detectors such as \acl{aLIGO} \citep[\acsu{aLIGO};][]{2015CQGra..32g4001T}, \acl{AdV} \citep[\acsu{AdV};][]{TheVirgo:2014hva} and KAGRA \citep{Aso:2013eba}. Searches of data from the first observing run \citep[][]{Aasi:2013wya} of \ac{aLIGO} yielded three likely \ac{BBH} coalescences: GW150914 \citep{Abbott:2016blz}, GW151226 \citep{Abbott:2016nmj} and LVT151012 \citep{TheLIGOScientific:2016qqj,TheLIGOScientific:2016pea}. \change{A further \ac{BBH} coalescence, GW170104, has been reported from the on-going second observing run \citep{Abbott:2017vtc}}. \ac{GW} observations give a unique insight into the properties of \acp{BBH}.  We will examine one of the ways in which black hole spin measurements can be used to constrain formation mechanisms.

\ac{GW} observations inform our understanding of \ac{BBH} evolution in two ways: from the merger rate, and from the properties of the individual systems.
The merger rate of \acp{BBH} is inferred from the number of detections; it is uncertain as a consequence of the small number of \ac{BBH} observations so far. \change{Currently}, merger rates are estimated to be \change{$12$--$213~\mathrm{Gpc^{-3}\,yr^{-1}}$ \citep{Abbott:2016nhf,Abbott:2017vtc}}. These rates are broadly consistent with predictions from both population synthesis models of isolated binary evolution \citep[e.g.,][]{2003MNRAS.342.1169V,Lipunov:2009,Dominik:2012kk,lrr-2006-6} and dynamical formation models \citep[e.g.,][]{Sigurdsson:1993,2008ApJ...676.1162S,2015PhRvL.115e1101R}. Possible progenitors systems of \acp{BBH}, including Cyg~X-3 \citep{Belczynski:2012jc}, IC~10~X-1 \citep{Bulik:2008ab} and NGC~300~X-1 \citep{Crowther01032010} provide some additional limits on \ac{BBH} merger rates, but extrapolation is hindered by current observational uncertainties.  

The parameters of individual systems can be estimated by comparing the measured \ac{GW} signal with template waveforms \citep{TheLIGOScientific:2016wfe}. The masses and spins of the \acp{BH} can be measured through their influence on the inspiral, merger and ringdown of the system \citep{Cutler:1994ys,1995PhRvD..52..848P,PhysRevD.47.2198}.
The distribution of parameters observed by \ac{aLIGO} will encode information about the population of \acp{BBH}, and may also help to shed light on their formation channels \citep{O'Shaughnessy:2006wh,2041-8205-725-1-L91,PhysRevD.87.104028,PhysRevD.89.124025, 2015MNRAS.450L..85M,2015arXiv150407802S,2017MNRAS.465.3254M,Vitale:2015tea}

Stellar-mass \acp{BH} are expected to be born spinning, with observations suggesting their dimensionless spin parameters $\chi$ take the full range of allowed values between $0$ and $1$ \citep{2011CQGra..28k4009M,Fragos:2014cva,2015PhR...548....1M}. 
Stars formed in binaries are expected to have their rotational axis aligned with the orbital angular momentum \citep[e.g.,][]{1988ComAp..12..169B,Albrecht:2007rf}, although there is observational evidence this is not always the case \citep[e.g.,][]{2009Natur.461..373A,2014ApJ...785...83A}. 
Even if binaries are born with misaligned spins, there are many processes in binary evolution which can act to align the spin of stars, such as realignment during a stable mass accretion phase \citep{1975ApJ...195L..65B,2000ApJ...541..319K,2005MNRAS.363...49K}, accretion onto a \ac{BH} passing through a \ac{CE} event \citep{2013A&ARv..21...59I}, and realignment through tidal interactions in close binaries \citep[e.g.,][]{1977A&A....57..383Z,1981A&A....99..126H}. 

On the other hand, asymmetric mass loss during supernova explosions can tilt the orbital plane in binaries \citep{1995MNRAS.274..461B,2000ApJ...541..319K}, leading to \ac{BH} spins being misaligned with respect to the orbital angular momentum vector. Population synthesis studies of \ac{BH} X-ray binaries predict that these misalignments are generally small \citep{2000ApJ...541..319K}, with \citet{2010ApJ...719L..79F} finding that the primary \ac{BH} is typically misaligned by $\lesssim 10 \degree$.  However, electromagnetic observations of high mass X-ray binaries containing \acp{BH} have hinted that the \acp{BH} may be more significantly misaligned \citep{2015PhR...548....1M}. One such system is the microquasar V4641 Sgr \citep{2001ApJ...555..489O,2008MNRAS.391L..15M} where the primary \ac{BH} is has been interpreted to be misaligned by $> 55\degree$.   

Alternatively, \acp{BBH} can form dynamically in dense stellar environments such as globular clusters. In these environments, it is expected that the distribution of \ac{BBH} spin--orbit misalignment angles is isotropic \citep[e.g.,][]{2016ApJ...832L...2R}. The distribution of \ac{BBH} spin--orbit misalignments therefore contains information about their formation mechanisms.  

Constraints on spin alignment from \ac{GW} observations so far are weak \citep{TheLIGOScientific:2016wfe,Abbott:2016izl,TheLIGOScientific:2016pea,Abbott:2017vtc}. Some configurations, such as anti-aligned spins for GW151226 \citep{Abbott:2016nmj}, are disfavoured; however, there is considerable uncertainty in the spin magnitude and orientation. Determining the spins precisely is difficult because their effects on the waveform can be intrinsically small (especially if the the source is viewed face on), and because of degeneracies between the spin and mass parameters \citep{1995PhRvD..52..848P,Baird:2012cu,Farr:2015lna}. Although the spins of individual systems are difficult to measure, here we show it is possible to use inferences from multiple systems to build a statistical model for the population \citep[cf.][]{Vitale:2015tea}.


This paper describes how to combine posterior probability density functions on spin--orbit misalignment angles from multiple \ac{GW} events to explore the underlying population.

\change{We develop a hierarchical analysis in order to combine multiple \ac{GW} observations of \ac{BBH} spin--orbit misalignments to give constraints on the fractions of \acp{BBH} forming through different channels. We consider different populations of potential spin--orbit misalignments, each representing different assumptions about binary formation, and use the \ac{GW} observations to infer the fraction of binaries from each population. In the field of exoplanets, similar hierarchical analyses have been used to make inference on the frequency of Earth-like exoplanets from measurements of the period and radius of individual exoplanet candidates \citep[e.g.,][]{2014ApJ...795...64F,Farr:2014}.  Other examples of the use of hierarchical analyses in astrophysics include modelling a population of trans-Neptunian objects \citep{Loredo:2004}, measurements of spin--orbit misalignments in exoplanets \citep{2012ApJ...754L..36N}, measurements of the eccentricity distribution of exoplanets \citep{2010ApJ...725.2166H} and the measurement of the mass distribution of galaxy clusters \citep{Lieu:2017xkq}.}

In Section~\ref{sec:models} we introduce our simplified population synthesis models for \acp{BBH}, paying special attention to the \ac{BH} spins. We briefly describe in Section~\ref{sec:pe} the \ac{PE} pipeline that will be employed to infer the properties (such as misalignment angles) of real \ac{GW} events, and discuss previous spin-misalignment studies in the literature. We introduce a framework for combining posterior probability density functions on spin--orbit misalignment angles from multiple \ac{GW} events to explore the underlying population in Section~\ref{sec:hierarchical}. We demonstrate the method using a set of mock \ac{GW} events in Section~\ref{sec:results}, and show that tens of observations will be sufficient to distinguish subpopulations of coalescing binary black holes, \change{assuming spin magnitudes of $\sim 0.7$.  Lower typical \ac{BH} spin magnitudes would reduce the accuracy with which the spin-orbit misalignment angle can be measured, therefore requiring more observations to extract information about subpopulations.}  We also show that more extreme models, such as the hypothesis that all \acp{BBH} have their spins exactly aligned with the orbital angular momentum, can be ruled out at a $5~\sigma$ confidence level with only $\mathcal{O}(5)$ observations of rapidly spinning \acp{BBH}. Finally, we conclude and suggest areas which require further study in Section~\ref{sec:conclusions}.

\section{BH spin misalignment models}
\label{sec:models}

Owing to the many uncertainties pertaining to stellar spins and their evolution in a binary, and the fact that keeping track of stellar spin vectors can be computationally intensive, many population synthesis models choose not to include spin evolution. However, the distribution of spins of the final merging \acp{BH} is one of the observables that can be measured with the advanced \ac{GW} detectors. In this section, we therefore implement a simplified population synthesis model to evolve an ensemble of binaries that will be detectable with \ac{aLIGO} and \ac{AdV} and predict their distributions of spin--orbit misalignments.  We describe the assumed mass distribution, spin distribution, and spin evolution below.

\subsection{Mass and spin magnitude distribution}

We assume the same simplified mass distribution for all of our models, so that any differences in the final spin distributions are purely due to our assumptions about the spin--orbit misalignments described in the next section.  There are many uncertainties in the evolution of isolated massive binaries, including (but not limited to) uncertainties in the initial distributions of the orbital elements \citep{deMink:2015yea}, the strength of stellar winds in massive stars \citep{2010ApJ...714.1217B}, the effect of rotation of massive stars on stellar evolution \citep{2013ApJ...764..166D,2015arXiv150702286R}, the natal kicks (if any) given to \acp{BH} \citep{2012MNRAS.425.2799R,2016MNRAS.456..578M,Mirabel:2016,Repetto:2017} and the efficiency of the \ac{CE} \citep{2013A&ARv..21...59I,2016A&A...596A..58K}. Population synthesis methods are large Monte-Carlo simulations using semi-analytic prescriptions in order to explore the effect these uncertainties have on the predicted distributions of compact binaries.  Instead, we adopt a number of simplifications that allow us to produce an astrophysically plausible distribution which should not, however, be considered representative of the actual mass distribution of \acp{BBH}.

We simulate massive binaries with semimajor axis $a$ drawn from a distribution uniform in $\ln a$ \citep{1983ARA&A..21..343A}. The components of the binary are a massive primary \ac{BH} with mass $m_1$ and a secondary star at the end of its main sequence lifetime.

The primary \ac{BH} was formed from a massive star with \ac{ZAMS} mass $m_1^\mathrm{ZAMS}$ drawn from the \ac{IMF} with a power law index of $-2.35$ \citep{1955ApJ...121..161S,2001MNRAS.322..231K}. The mass ratio of the binary at \ac{ZAMS} $q^\mathrm{ZAMS}$ is drawn from a flat distribution $[0,1]$. The mass of the secondary star is given by $m_{2}^\mathrm{ZAMS} = q^\mathrm{ZAMS}  m_{1}^\mathrm{ZAMS}$.

We calculate the final remnant mass $m_{i}$ as a function of the \ac{ZAMS} mass $m_{i}^\mathrm{ZAMS}$ for each star using a fit to Figure~12 in \citet{RevModPhys.74.1015}. For stars with $30 < m_{i}^\mathrm{ZAMS} / M_\odot < 50$, in which range \acp{BH} are formed after some delay by fall-back of ejecta, we use
\begin{equation}
\label{eq:mf}
m_{i} = 30 \left( \frac{m_{i}^\mathrm{ZAMS}}{50 M_\odot} \right)^{\alpha} M_\odot,
\end{equation} 
with $\alpha = 3.9$. For more massive stars with $m_{i}^\mathrm{ZAMS} > 50 M_\odot$, which are massive enough to directly collapse during the iron-core collapse to form \acp{BH}, we use
\begin{equation}
\label{eq:mfabove50}
m_{i} = 0.6  m_{i}^\mathrm{ZAMS}. 
\end{equation}
We only consider \acp{BBH} with component masses above $10 M_\odot$ below, consistent with \ac{aLIGO} detections to date \change{\citep{TheLIGOScientific:2016pea,Abbott:2017vtc}}, and hence omit stars with \ac{ZAMS} masses below $30 M_\odot$ from our population.

We assume that the binary has negligible eccentricity $e = 0$, appropriate for post-\ac{CE} systems. In all models we have assumed that both main-sequence stars in the binary are born with their rotation axis aligned with the orbital angular momentum axis. In general, the first supernova will misalign the spins due to any natal kick imparted on the remnant. There are expected to be mass-transfer phases between the first and second supernovae which may realign both the spins of the primary \ac{BH} and the secondary \ac{BH} progenitor; we vary the assumed degree of realignment in our models.

We assume that \acp{BH} receive natal kicks comparable to those received by neutron stars \citep{2005MNRAS.360..974H}, namely drawn from a Maxwellian with a root-mean-square velocity of $\sim~250~\mathrm{km\,s}^{-1}$. This assumption will lead to the maximum amount of spin misalignment, and may be consistent with neutrino-driven kicks; if the natal kicks are due to asymmetric ejection of baryonic matter, then any fall-back \citep{Fryer:2011cx} onto \acp{BH} during formation will reduce the kick magnitude and thus the spin misalignment.

\ac{BH} spins magnitudes can take any value $0 \leq \chi_i < 1$, but we set $\chi_i = 0.7$ for all our \acp{BH}. High spin magnitudes are consistent with measurements from X-ray observations \citep[cf.][]{2015PhR...548....1M}, and lie toward the upper end of the range allowed by current \ac{GW} observations \change{\citep{TheLIGOScientific:2016pea,Abbott:2017vtc}}. Such spins are large enough to ensure spin effects on the gravitational waveform are significant, providing an opportunity for us to demonstrate our hierarchical approach, but small enough that we do not have to worry about the validity of the model gravitational waveform. Uncertainties in the relationship between pre-supernova stellar spins and \ac{BH} spins mean it is not currently possible to produce a realistic distribution of spins from first principles, although a direct translation is often assumed, e.g., by \citet{Kushnir:2016}. If the distribution of \ac{BH} spin magnitudes in nature favours smaller values, then more observations will be required to draw the conclusions we find here. The methodology we use here can be extended to models including \ac{BH} spin magnitudes, which could potentially give us further information regarding formation mechanisms.

After a supernova, we establish whether the binary remains bound and, for those that do, find the new orbital elements \citep{1961BAN....15..265B,1995MNRAS.274..461B,1996ApJ...471..352K}. Of the remaining bound systems, we are only interested in those binaries which merge due to the emission of gravitational radiation within a Hubble time, as these are the binaries that are potentially observable with \ac{GW} detectors.

\subsection{Models for spin--orbit misalignment distributions}

We model the overall population of \acp{BBH} as a mixture of $4$ subpopulations, each of which makes differing assumptions leading to distinct spin--orbit misalignment distributions.  

We define the spin--orbit misalignment angle as the angle between the spin vector $\uvect{S}_i$ of binary component $i \in 1,2$ and the (Newtonian) orbital angular momentum vector $\uvect{L}$,
\begin{equation}
\cos{\theta_i} = \uvect{S}_i \cdot \uvect{L},
\end{equation}
where
\begin{equation}
\vect{S}_i = \chi_{i} m_{i}^{2} \uvect{S}_i,
\end{equation}
and $m_i$ is the component mass ($m_1 \geq m_2$).\footnote{Throughout this paper we use geometric units $G = c = 1$ unless otherwise stated.} 
We will consider how a set of spin-misalignment measurements could be used to infer \ac{BBH} formation mechanisms.

\begin{figure*}
\includegraphics[width = 0.3\textwidth]{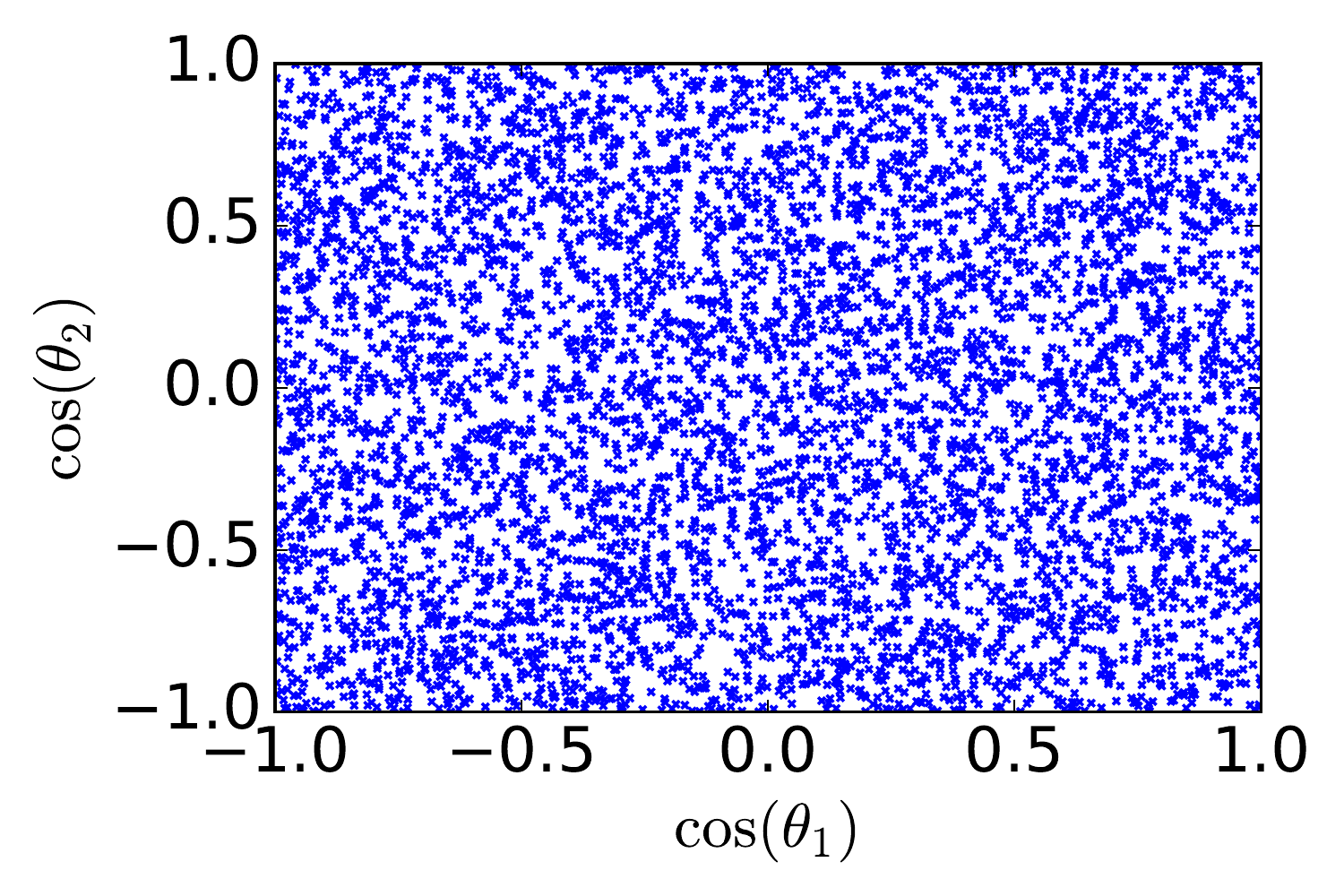}
\quad
\includegraphics[width = 0.3\textwidth]{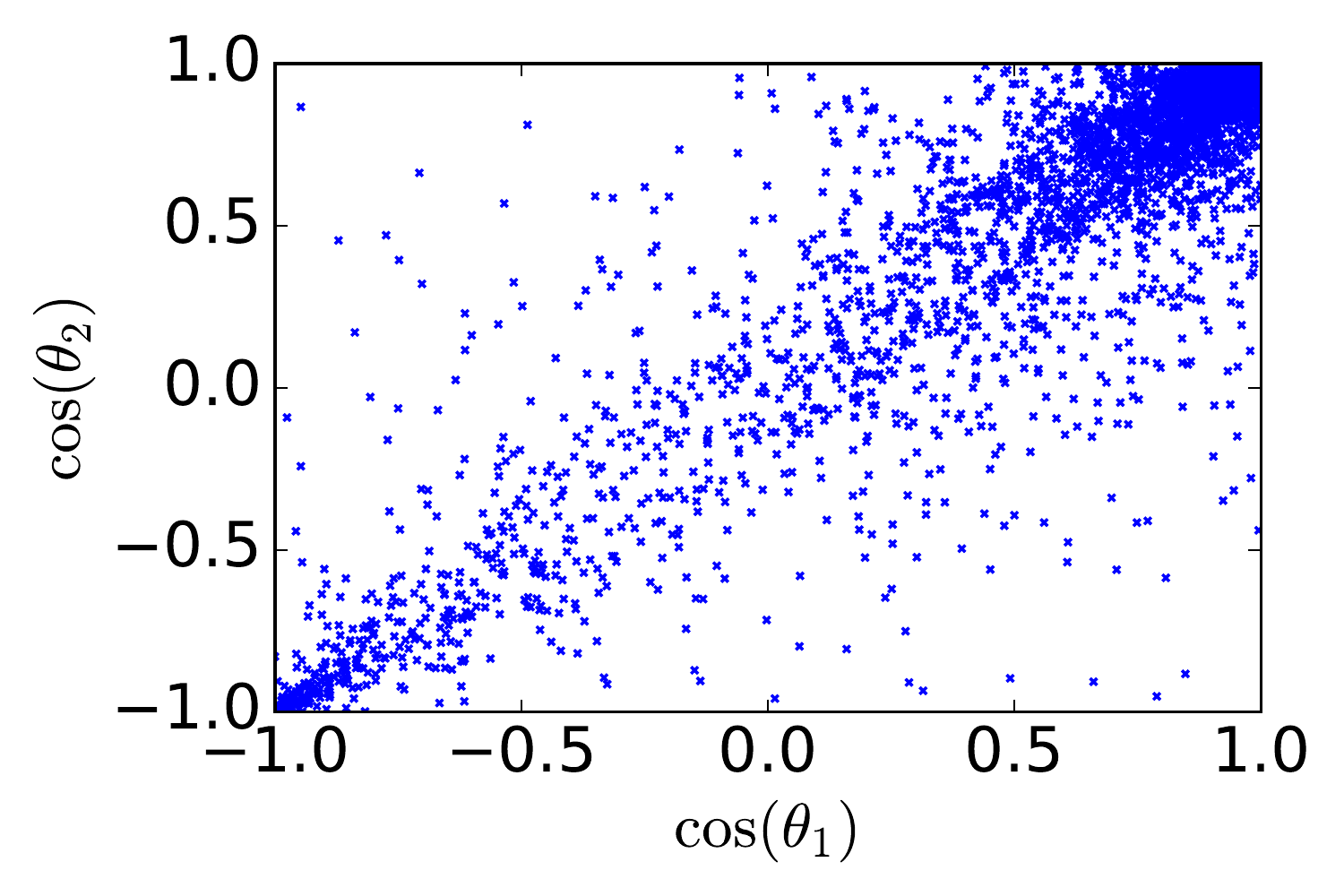}
\quad
\includegraphics[width = 0.3\textwidth]{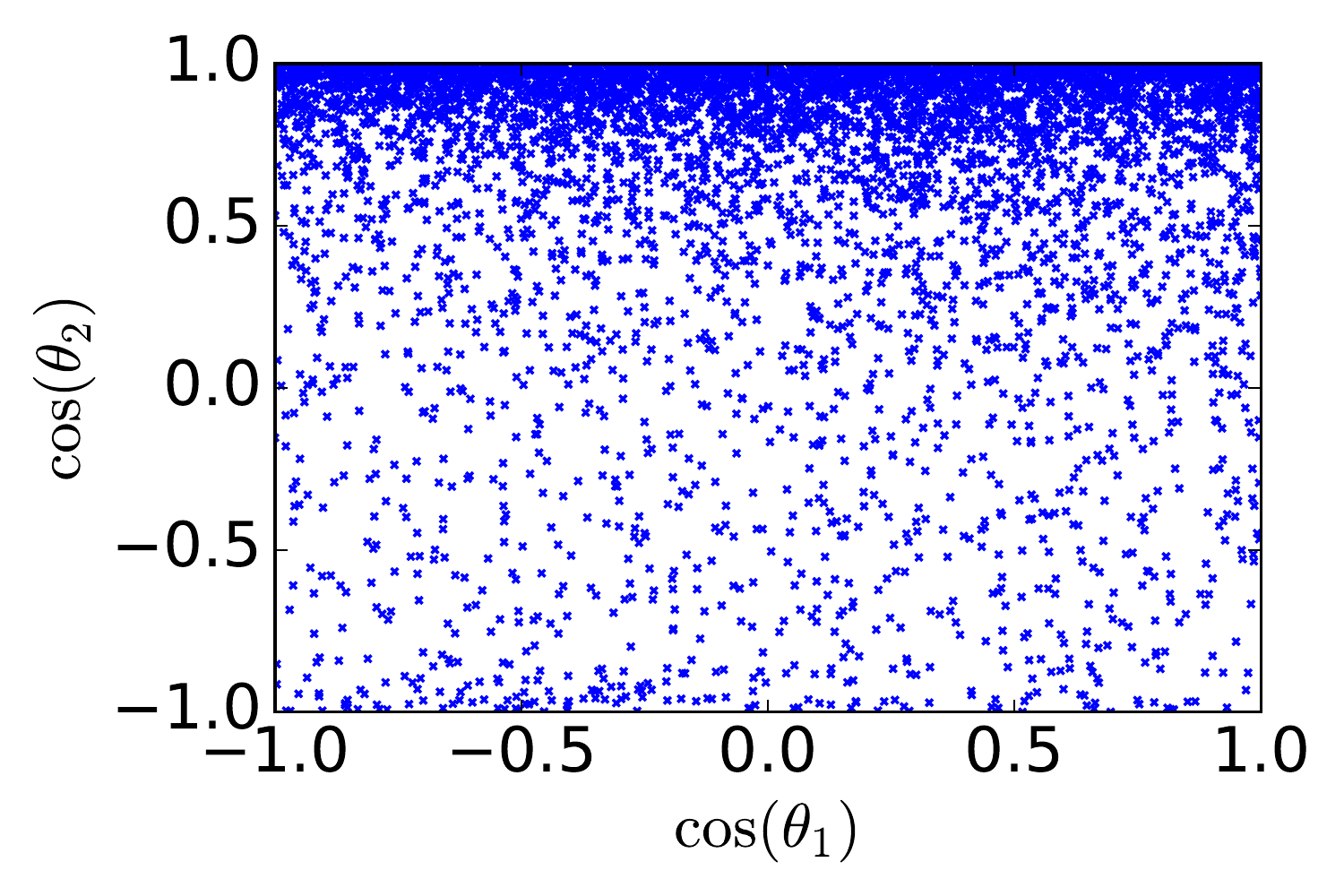}
\caption{Three of the four astrophysically motivated subpopulations making up our mixture model for \ac{BBH} spin misalignment angles $\theta_1$ and $\theta_2$ described in Section~\ref{sec:models}. Subpopulation~1 (not shown) has both spins perfectly aligned ($\cos \theta_1 = \cos \theta_2 = 1$), so all points would lie in the top right corner. In subpopulation~2 (left),  both spins are drawn from an isotropic distribution, and so the samples are distributed uniformly in the plane. In subpopulation~3 \ac{BH} spins are aligned with the orbital angular momentum just prior to the second supernova. In subpopulation~4, the secondary \ac{BH} has its spin aligned with the orbital angular momentum prior to the second supernova, whilst the primary is misaligned. See Section~\ref{sec:models} for more details. \change{Spins are quoted at a \ac{GW} frequency of $f_\mathrm{ref} = 10~\mathrm{Hz}$}.}
\label{fig:models}
\end{figure*}

\begin{description}

  \item[\textbf{Subpopulation 1: Exactly aligned}] We assume that irrespective of all prior processes, both \acp{BH} have their spins aligned with the orbital angular momentum vector after the second supernova, such that $\cos{\theta_1} = \cos{\theta_2} = 1$. This may be the case if \acp{BH} receive no kicks. \ac{GW} searches often assume \acp{BBH} have aligned spins as this simplification makes the search less computationally demanding \citep{TheLIGOScientific:2016qqj}.
  
  \item[\textbf{Subpopulation 2: Isotropic/dynamical formation}] We assume that \acp{BBH} are formed dynamically, such that the distribution of spin angles is isotropic. Initially isotropic distributions of spins remain isotropic \citep{PhysRevD.70.124020}. We still generate the binary mass distribution with our standard approach, so that the only difference in \acp{BBH} between this model and the others is the spin distribution.
  
  \item[\textbf{Subpopulation 3: Alignment before second SN}] Motivated by \citet{2000ApJ...541..319K}, we assume that the spins of both components are aligned with the orbital angular momentum after a \ac{CE} event and prior to the second supernova. The tilt of the orbital plane caused by the second supernova is then taken to be the spin misalignment angle of \textit{both} components, i.e.\ $\cos{\theta_1} = \cos{\theta_2}$. As we discuss in Section~\ref{subsec:precession}, these spins freely precess from the time of the second supernova up until merger. This precession somewhat scatters these angles, but leaves them with generally similar values, as seen in Figure~\ref{fig:models}.
  
  \item[\textbf{Subpopulation 4: Alignment of secondary}] We follow the standard mass-ratio model with effective tides presented in \citet{PhysRevD.87.104028}, which assumes that after the first supernova, the secondary is realigned via tides or the \ac{CE} prior to the second supernova. However, the primary \ac{BH} is not realigned.  Because the binary's orbit shrinks during \ac{CE} ejection, the kick velocity of the secondary is small relative to its pre-supernova velocity, causing the secondary to be only mildly misaligned (in general $\theta_1 > \theta_2$).
\end{description}

\begin{figure}
\centering
\includegraphics[width=0.45\textwidth]{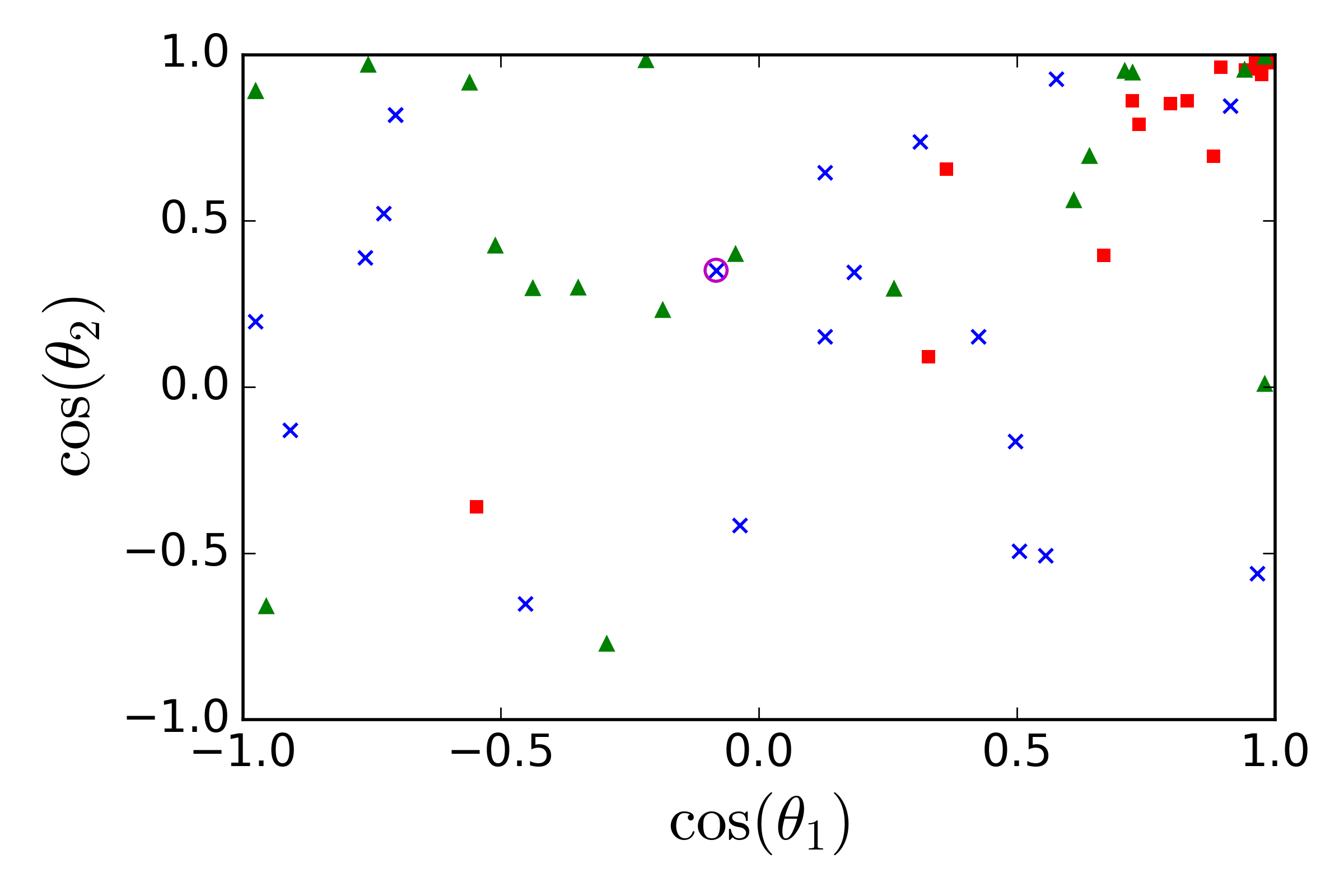}
\caption{True values of \ac{BH} spin--orbit misalignment angles $\cos{\theta_{1}}$ and $\cos{\theta_{2}}$ for a mixture of $20$ draws from each of our four subpopulations. Exactly aligned systems from subpopulation~1 sit in the upper right corner of this diagram and thus are not shown. Systems drawn from subpopulation~2 are shown as blue crosses, those from subpopulation~3 as red squares and those from subpopulation~4 as green triangles. \change{The injection plotted in Figures~\ref{fig:typical_event_posterior_costheta1_costheta2} and \ref{fig:typical_event_fake_posterior_costheta1_costheta2} is circled in magenta.}}
\label{fig:injections}
\end{figure}
%


We generate several thousand samples from each of these models.  We plot samples from the four subpopulations in the $\{\cos{\theta_1}, \cos{\theta_2}\}$ plane of the misalignment angles in Figure~\ref{fig:models}. From each of these models we randomly select $20$ mock detected systems (for a total of $N = 80$) with component masses between $10 M_\odot$ and $40 M_\odot$ \citep[cf.][]{TheLIGOScientific:2016pea}; we describe the analysis of these mock \ac{GW} signals in Section~\ref{sec:pe}. We show the true values of the spin misalignment angles for the mock detected systems in Figure~\ref{fig:injections}.

We use a \change{roughly} astrophysical distribution of systems with sky positions and inclinations randomly chosen, and distances $D_\mathrm{L}$ distributed uniformly in volume, $p(D_\mathrm{L}) \propto D_\mathrm{L}^{2}$, such that the distribution of \ac{SNR} $\rho$ is $p(\rho) \propto \rho^{-4}$ \citep{2011CQGra..28l5023S}.  \change{This approximates the Universe as being static, with constant merger rates per unit time, at the distance scales probed by current detectors.}  We use a detection threshold (minimum network \ac{SNR}) of $\rho_\mathrm{min} = 12$ \citep{Aasi:2013wya,2015ApJ...804..114B}.  

\subsection{Precession and spin--orbit resonances}
\label{subsec:precession}

After the second supernova, the evolution of the \ac{BBH} is purely driven by relativistic effects and the orbit decays due to the emission of gravitational radiation \citep{PhysRev.131.435,Peters:1964}. As the \acp{BH} orbit, their spins precess around the total angular momentum \citep{1994PhRvD..49.6274A,2014LRR....17....2B}. In order to predict the spin misalignment angles when the frequency of \acp{GW} emitted by the binary are high enough (or equivalently when the orbital separation of the binary is sufficiently small) to be in the \ac{aLIGO} band ($f_\mathrm{GW} > 10~\mathrm{Hz}$), we take into account the \ac{PN} evolution of the spins by evolving the ten coupled differential equations given by Equations (14)--(17) in \citet{PhysRevD.87.104028}. We begin our integrations at an orbital separation $a = 1000 M$, and integrate up until $f_\mathrm{GW} = 10~\mathrm{Hz}$.\footnote{A more efficient method of evolving binaries from wide orbital separations to the frequencies where they enter the \ac{aLIGO} band was introduced in \citet{2015PhRvL.114h1103K} and \citet{2015arXiv150603492G}. This exploits the hierarchy of timescales in the problem and integrates precession averaged equations of motion on the radiation reaction timescale, rather than integrating the orbit-averaged equations we use here.}  

Some of these binaries are attracted to spin--orbit resonances \citep{PhysRevD.70.124020}. \change{In particular, the binaries from subpopulation~4 are attracted to the $\Delta \Phi = \pm 180\degree$ resonance}, where $\Delta \Phi$ is the angle between the projection of the two spins on the orbital plane. The \change{current generation of} ground-based \ac{GW} observations are generally insensitive to this angle \change{for binary black holes \citep{2015PhRvD..91b4043S,Abbott:2016izl},} \change{and the waveform model we use does not include it}, so we focus on distinguishing subpopulations through the better-measured $\theta_1$ and $\theta_2$ angles. 

\section{\ac{GW} parameter estimation}
\label{sec:pe}

\subsection{Signal analysis and inference}

The strain measured by a \ac{GW} detector is a combination of detector noise and (possibly) a \ac{GW} signal $h(\vect{\Theta}, t)$,
\begin{equation}
d(t) = n(t) + h(\vect{\Theta}, t).
\end{equation}
Here $\vect{\Theta}$ is the vector of parameters describing the \ac{GW} signal; for a general spinning circular \ac{BBH}, there are $15$ parameters.\footnote{These parameters are \citep[e.g.,][]{Veitch:2014wba}: two component masses $\{m_i\}$; six spin parameters describing $\{\vect{S}_i\}$; two sky coordinates; distance $D\sub{L}$; inclination and polarization angles; a reference time, and the orbital phase at this time.} Given a data stream, we want to infer the most probable set of parameters for that data. To estimate the properties of the signal, waveform templates are matched to the data \citep{Cutler:1994ys,Veitch:2014wba,TheLIGOScientific:2016wfe}. 

The posterior probability for the parameters is given by Bayes' theorem,
\begin{equation}
p(\vect{\Theta} | d) = \frac{p(d | \vect{\Theta}) p(\vect{\Theta})}{p(d)} ,
\label{eq:bayestheorem}
\end{equation}
where $p(d | \vect{\Theta})$ is the likelihood of observing the data given a choice of parameters, $p(\vect{\Theta})$ is the prior on those parameters, and the evidence $p(d)$ is a normalisation constant for the purposes of \ac{PE}. The prior encodes our belief about the parameters before we considered the data: we assume that sources are uniformly distributed across the sky and in volume; that spin magnitudes are uniformly distributed between $0$ and $1$; that spin orientations and the binary orientation are uniformly distributed across the surface of the sphere, and that component masses are uniformly distributed up to a maximum of $150 M_\odot$ \citep[cf.][]{TheLIGOScientific:2016wfe}.
The likelihood is calculated from the residuals between the data and the signal template, assuming that the noise is Gaussian \citep{Cutler:1994ys}:
\begin{equation}
p(d | \vect{\Theta}) \propto \exp \left[ -\frac{1}{2} \left( d - h(\vect{\Theta}) | d - h(\vect{\Theta}) \right) \right] ,
\end{equation}
where the inner product $\left(g | h \right)$ is given by \citep{Finn:1992wt}
\begin{equation}
\left( g \middle| h \right) = 4 \Re \int_{f_{\mathrm{low}}}^{\infty} \frac{\tilde{g}(f)
\tilde{h}^{*}(f)}{S_n (f)} \mathrm{d}f ,
\end{equation}
and $S_n (f)$ is the (one-sided) noise power spectral density \citep{2015CQGra..32a5014M}, which we take to be the design sensitivities for \ac{aLIGO} and \ac{AdV} respectively, with $f_\mathrm{low} = 10~\mathrm{Hz}$ as is appropriate for the advanced detectors.

\begin{figure}
\centering
\includegraphics[width=0.45\textwidth]{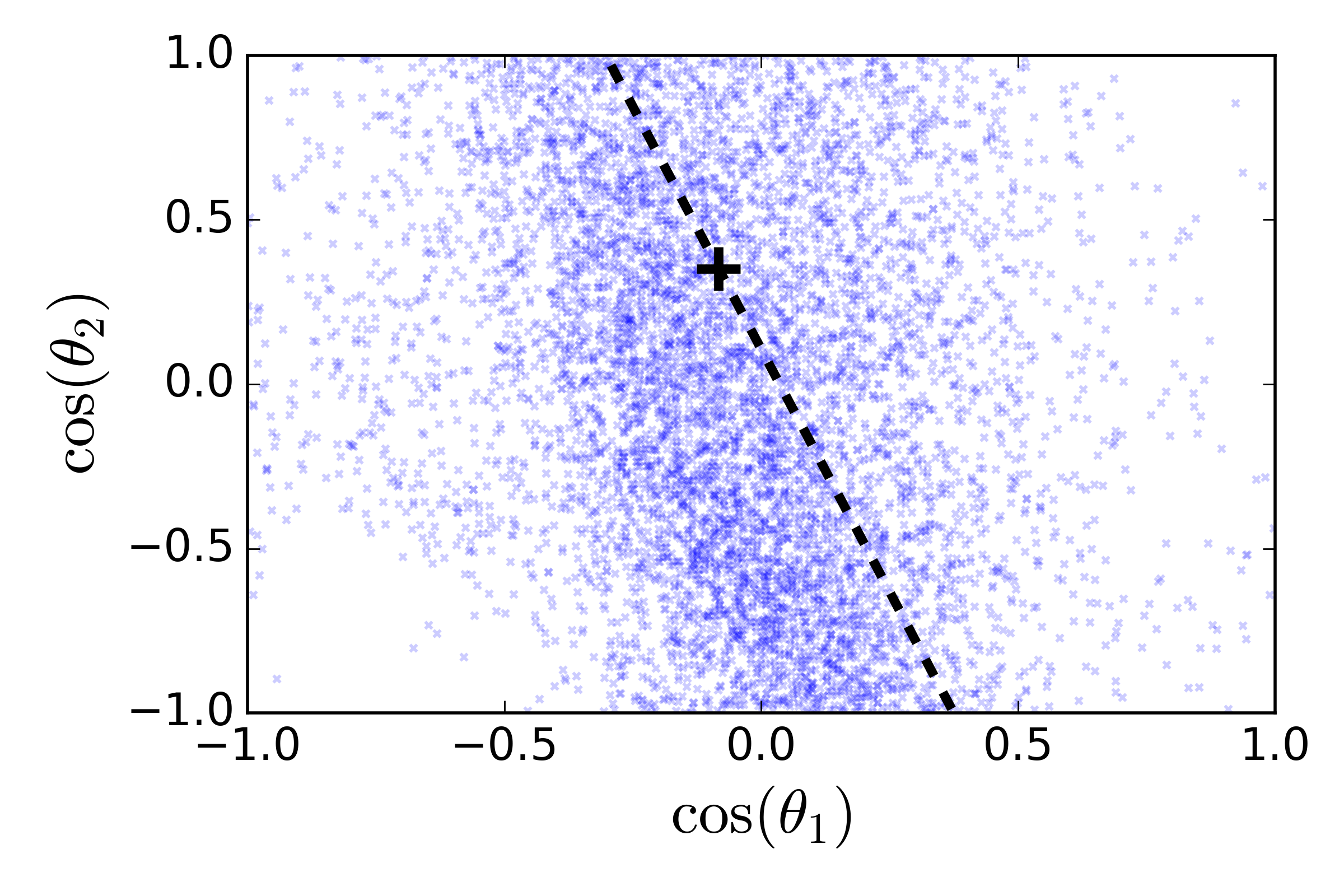} 
\caption{Marginalised posterior samples for one of the $80$ events shown in Figure~\ref{fig:injections}, generated by analysing mock \ac{GW} data using \texttt{LALInference}. The true spin--orbit misalignments (thick black plus) for this event were $\cos{\theta_1} = -0.08$ and $\cos{\theta_2} = 0.35$, with a network SNR of $15.35$. The dashed black diagonal line shows the line of constant $\chi_\mathrm{eff}$.}
\label{fig:typical_event_posterior_costheta1_costheta2}
\end{figure}
%


We sample the posterior distribution using the publicly available, Bayesian \ac{PE} code \texttt{LALInference} \citep{Veitch:2014wba}.\footnote{Available as part of the LIGO Scientific Collaboration Algorithm Library (LAL) \href{https://wiki.ligo.org/DASWG/LALSuite}{https://wiki.ligo.org/DASWG/LALSuite}.} 
For each event we obtain $\nu \sim 5000$ independent posterior samples. We show an example of the marginalised posterior distribution in $\{\cos{\theta_1}, \cos{\theta_2}\}$ space for one of our 80 events in Figure~\ref{fig:typical_event_posterior_costheta1_costheta2}. \change{Unless otherwise stated, we quote all parameters at a reference frequency $f_\mathrm{ref} = 10~\mathrm{Hz}$.}

We sample in the system frame \citep{2014PhRvD..90b4018F}, where the binary is parametrised by the masses and spin magnitudes of the two component \acp{BH} $\{m_i\}$ and $\{\chi_i\}$, 
the spin misalignment angles $\{\theta_i\}$, the angle $\Delta \Phi$ between the projections of the two spins on the orbital plane, and the angle $\beta$ between the total and orbital angular momentum vectors. We find, in agreement with similar studies such as \citet{2015arXiv150303179L} and \citet{2015arXiv150606032M}, that there is a strong preference for detecting \acp{GW} from 
nearly face-on binaries, since \ac{GW} emission is strongest perpendicular to the orbital plane.

Following common practice in \ac{PE} studies, we use a special realisation of Gaussian noise which is exactly zero in each frequency bin \citep{2015arXiv150305953V,2015arXiv150705587T}. Real \ac{GW} detector noise will be non-Gaussian and non-stationary, and events will be recovered with non-zero noise.\footnote{Non-stationary, non-Gaussian noise has been shown not to affect average \ac{PE} performance for binary neutron stars \citep{2015ApJ...804..114B}; however, these noise features could be more significant in analysing the shorter duration \ac{BBH} signals.} Non-zero noise-realisations will mean that in general the maximum likelihood parameters do not match the injection parameters; in the Gaussian limit, however, using a zero noise realisation is equivalent to averaging over a large number of random noise realisations, such that these offsets approximately cancel out \citep[cf.][]{Vallisneri:2011}. This assumption makes it straightforward to compare the posterior distributions, as differences only arise from the input parameters and not any the specific noise realisation.

\subsection{Previous studies}

\citet{Vitale:2015tea} study \ac{GW} measurements of \ac{BH} spin misalignments in compact binaries containing at least one \ac{BH}. They consider both \acp{BBH} (using \texttt{IMRPhenomPv2} waveforms as we do here) and \ac{NSBH} binaries (using inspiral-only \texttt{SpinTaylorT4} waveforms). They fit a mixture model allowing for both a preferentially aligned/anti-aligned component and an isotropically misaligned component, excluding aligned/anti-aligned systems. They find that $\sim100$ detections yield a $\sim10\%$ precision for the measured aligned fraction. One of the main limitations of the analysis performed by \citet{Vitale:2015tea} is that they only consider models which are mutually exclusive, \change{although this should not affect their results since the excluded region for their nonaligned model is negligible.} Here, all of our formation models overlap in the parameter space of spin--orbit misalignment angles. \change{Therefore, we cannot directly apply the formalism of \citet{Vitale:2015tea}}. The framework we develop here is able to correctly determine the relative contributions of multiple models, even when those models overlap in parameter space significantly, as expected in practice. 



There have also been significant advances in the past few years in the understanding of \ac{PN} spin--orbit resonances. These resonances occur when \ac{BH} spins become aligned or anti-aligned with one another and precess in a common plane around the total angular momentum \citep{PhysRevD.70.124020}. This causes binaries to be attracted to different points in parameter space identified by $\Delta \Phi$, the angle between the projections of the two \ac{BH} spins onto the orbital plane. \citet{2010PhRvD..81h4054K} have shown that these resonances are effective at capturing binaries with mass ratios $0.4 < q < 1$ and spins $\chi_i > 0.5$. For equal-mass binaries, \change{spin morphologies remain locked with binaries trapped in or out of resonance \citep{2017CQGra..34f4004G}}; however, it is unlikely for astrophysical formation scenarios to produce exactly equal mass binaries, although \citet{Marchant:2016} predict nearly equal masses for the chemically homogeneous evolution channel. 

\citet{PhysRevD.87.104028} show how the family of resonances that \acp{BBH} are attracted to can act as a diagnostic of the formation scenario for those binaries. \citet{2015arXiv150705587T} demonstrate that \ac{GW} measurements of spin misalignments can be used to distinguish between the two resonant families of $\Delta \Phi = 0\degree$ and $\Delta \Phi = \pm 180\degree$. They use a full \ac{PE} study to show that they can distinguish two families of \ac{PN} resonances. However, they only consider a small corner of parameter space which contains binaries which will become locked in these \ac{PN} resonances. 

Our study extends on those discussed in several ways:
\begin{enumerate}

\item Rather than focusing on specific systems preferred in previous studies, we use injections from an astrophysically motivated population. Our injections have total masses $M = m_1 + m_2$ in the range $10$--$40 M_\odot$ and an astrophysical distribution of \acp{SNR}.

\item The misalignment angles of our \acp{BH} are given by simple but astrophysically motivated models introduced in Section~\ref{sec:models}. 

\item For performing \ac{PE} on individual \ac{GW} events, we use the inspiral-merger-ringdown gravitational waveform \texttt{IMRPhenomPv2} model, rather than the inspiral-only waveforms used in some of the earlier studies.

\item Most importantly, we perform a hierarchical Bayesian analysis on the posterior probability density functions of a mock catalog of detected events in order to make inferences about the underlying population.

\end{enumerate}

\section{Hierarchical analysis for population inference}
\label{sec:hierarchical}

\ac{PE} on individual \ac{GW} events yields samples from the posterior distributions for parameters under astrophysical prior constraints.
We now wish to combine these individual measurements of \ac{BH} spin misalignment angles in order to learn about the underlying population, which may act as a diagnostic for binary formation channels and binary evolution scenarios. Importantly, we are able to do this without reanalysing the data for the individual events.


Given a set of reasonable population synthesis model predictions for \ac{BBH} spin misalignment angles, we would like to learn what mixture of those subpopulations best explains the observed data.   Here we assume that the subpopulation distributions representing different formation channels are known perfectly, and use the same subpopulations that we drew our injections from to set these distributions.  Thus, each subpopulation model $\Lambda_\ell$ ($\ell \in 1 \dots 4$) corresponds to a known distribution of source parameters $p(\vect{\Theta} | \Lambda_\ell)$. In practice, the uncertainty in the subpopulation models will be one of the challenges in carrying out accurate hierarchical inference.\footnote{The clustering approach of \citet{2015MNRAS.450L..85M,2017MNRAS.465.3254M}, which eschews assumptions about the subpopulation distributions, could provide an alternative pathway for robust but less informative inference on the data alone.}

The overall mixture model is described by hyperparameters $\lambda_\ell$, corresponding to the fraction of each of the four subpopupulations, such that
\begin{equation}
p(\vect{\Theta} | \vect{\lambda}) = \sum_{\ell = 1}^{4} \lambda_\ell p(\vect{\Theta} | \Lambda_\ell).
\end{equation}
We assume that each event comes from one of these subpopulations:
\begin{equation}
\sum_{\ell = 1}^{4} \lambda_\ell = 1,
\end{equation}
i.e.,\ $\vect{\lambda}$ is a unit simplex.

For any individual event $\alpha$ ($\alpha = 1,\ldots,N$) we have the posterior on $\vect{\Theta}$ given by
\begin{equation}
p(\vect{\Theta} | d_\alpha) = \frac{p(d_\alpha | \vect{\Theta}) p(\vect{\Theta})}{p(d_\alpha)},
\end{equation}
where  $p(\vect{\Theta})$ is the prior used by \texttt{LALInference}, $p(d_\alpha)$ is the evidence (which is only a normalising factor in our analysis), and we represent $p(\vect{\Theta} | d_\alpha)$ by a set of discrete samples $\{ \vect{\Theta}_i^{k}\}$ where $k = 1,\ldots,\nu_{\alpha}$.

We can write the likelihood for obtaining all of the events as the product over the individual likelihoods \citep{PhysRevD.81.084029,2010ApJ...725.2166H},
\begin{align}
p \left( \left\{ d_\alpha\right\}_{\alpha = 1}^{N} \middle| \vect{\lambda} \right) &= \prod_{\alpha = 1}^{N} p(d_\alpha | \vect{\lambda}) \\
&= \prod_{\alpha = 1}^{N} \int \mathrm{d} \vect{\Theta}_\alpha p(d_\alpha | \vect{\Theta}_\alpha) p(\vect{\Theta}_\alpha | \vect{\lambda}) \\
&= \prod_{\alpha=1}^{N} p(d_\alpha) \int \mathrm{d} \vect{\Theta}_\alpha \frac{p(\vect{\Theta}_\alpha | d_\alpha)}{p(\vect{\Theta}_\alpha)} p(\vect{\Theta}_\alpha | \vect{\lambda}) ,
\label{eq:marginalise_over_individual_events}
\end{align}
where we have marginalised over the physical parameters of the individual events, and used Bayes' theorem to obtain the final line. Since we have samples drawn from the posterior $p(\vect{\Theta}_\alpha | d_\alpha)$, we can approximate posterior-weighted integrals (posterior averages) as a sum over samples \citep[chapter 29]{MacKay2003},
\begin{equation}
\int \mathrm{d} \vect{\Theta}_\alpha p(\vect{\Theta}_\alpha | d_\alpha) f(\vect{\Theta}_\alpha) = \frac{1}{\nu_{\alpha}} \sum_{k=1}^{\nu_\alpha} f(\vect{\Theta}_{\alpha}^{k}),
\end{equation}
where $f(\vect{\Theta})$ is some general function. 
Thus, we can rewrite Equation~\eqref{eq:marginalise_over_individual_events} as
\begin{equation}
p \left( \left\{ d_\alpha \right\}_{\alpha=1}^{N} \middle| \vect{\lambda} \right) = \prod_{\alpha=1}^{N} \frac{{p(d_\alpha)}}{\nu_{\alpha}} \sum_{k=1}^{\nu_{\alpha}} \frac{p(\vect{\Theta}_\alpha^{\color{black}k} | \vect{\lambda})}{p(\vect{\Theta}_\alpha^{\color{black}k})}.
\label{eq:likelihood_sum_over_samples}
\end{equation}
In effect, for each event we reweigh the evidence calculated using our general \ac{PE} prior to what it would have been using a prior for the model of interest, and then combine these probabilities together to form a likelihood. 

\begin{figure}
\centering
\includegraphics[width=0.45\textwidth]{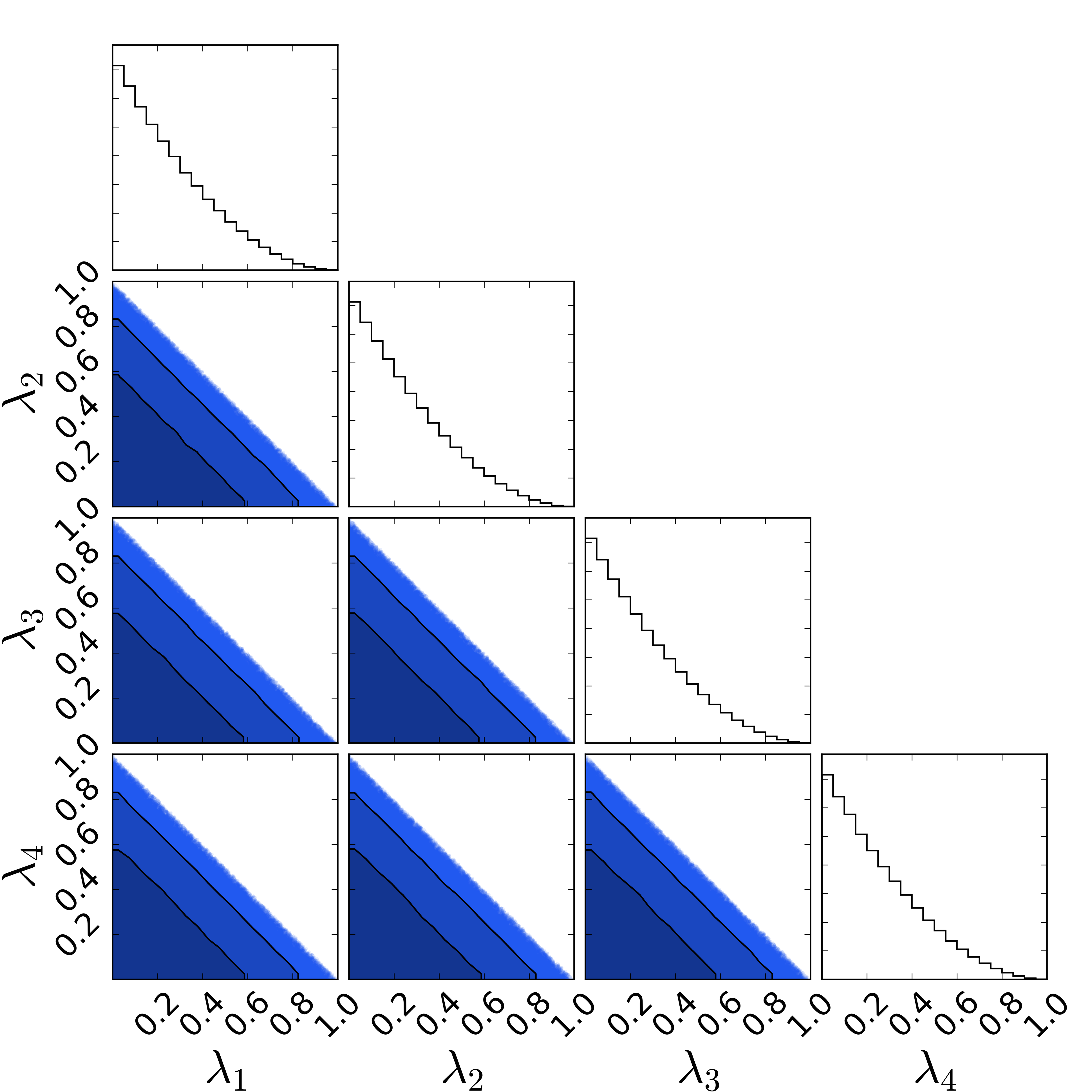}
\caption{Marginalised 1D and 2D probability density functions for the Dirichlet prior used for the analysis of the $\lambda$ parameters, which describe the fractional contribution of each of the four subpopulations introduced in Section~\ref{sec:models}. The constraint $\lambda_1 + \lambda_2 + \lambda_3 + \lambda_4 \equiv 1$ introduces correlations between parameters. \change{The shaded regions show the 68\% (darkest blue) and 95\% (middle blue) confidence regions, with the individual posterior samples outside these regions plotted as scatter points (lightest blue).}} 
\label{fig:dirichlet_prior}
\end{figure}

The posterior for $\vect{\lambda}$ is then
\begin{equation}
p \left( \vect{\lambda} \middle|  \left\{ d_\alpha \right\}_{\alpha=1}^{N}\right) \propto p \left( \left\{ d_\alpha \right\}_{\alpha=1}^{N} \middle| \vect{\lambda} \right) p(\vect{\lambda}),
\label{eq:posterior_lambda}
\end{equation}
for a choice of prior $p(\vect{\lambda})$. We assume a flat Dirichlet prior as shown in Figure~\ref{fig:dirichlet_prior}. We sample from this posterior on $\vect{\lambda}$ using \texttt{emcee} \citep{2013PASP..125..306F}, an affine-invariant ensemble sampler \citep{Goodman2010}.\footnote{Available from \href{http://dan.iel.fm/emcee/}{http://dan.iel.fm/emcee/}.} 

\section{Results}
\label{sec:results}

To gain a qualitative understanding of hierarchical modelling on the spin--orbit misalignment angles, we first consider inference under the assumption of perfect measurement accuracy for individual observations, and then introduce realistic measurement uncertainties.  We then analyse the scaling of the inference accuracy with the number of observations.
 
\subsection{Perfect measurement accuracy}

Here, we assume that \ac{aLIGO}--\ac{AdV} \ac{GW} observations could perfectly measure the spin--orbit misalignment angles of merging \acp{BBH}. In this case, the posterior is simply a delta function centered at the true value. Since our underlying astrophysical models have significant overlap in the $\{\cos(\theta_1)$, $\cos(\theta_2)\}$ plane, as shown in Figure~\ref{fig:models}, there is still ambiguity about which model a given event comes from. 

\begin{figure}
\centering
\includegraphics[width=0.45\textwidth]{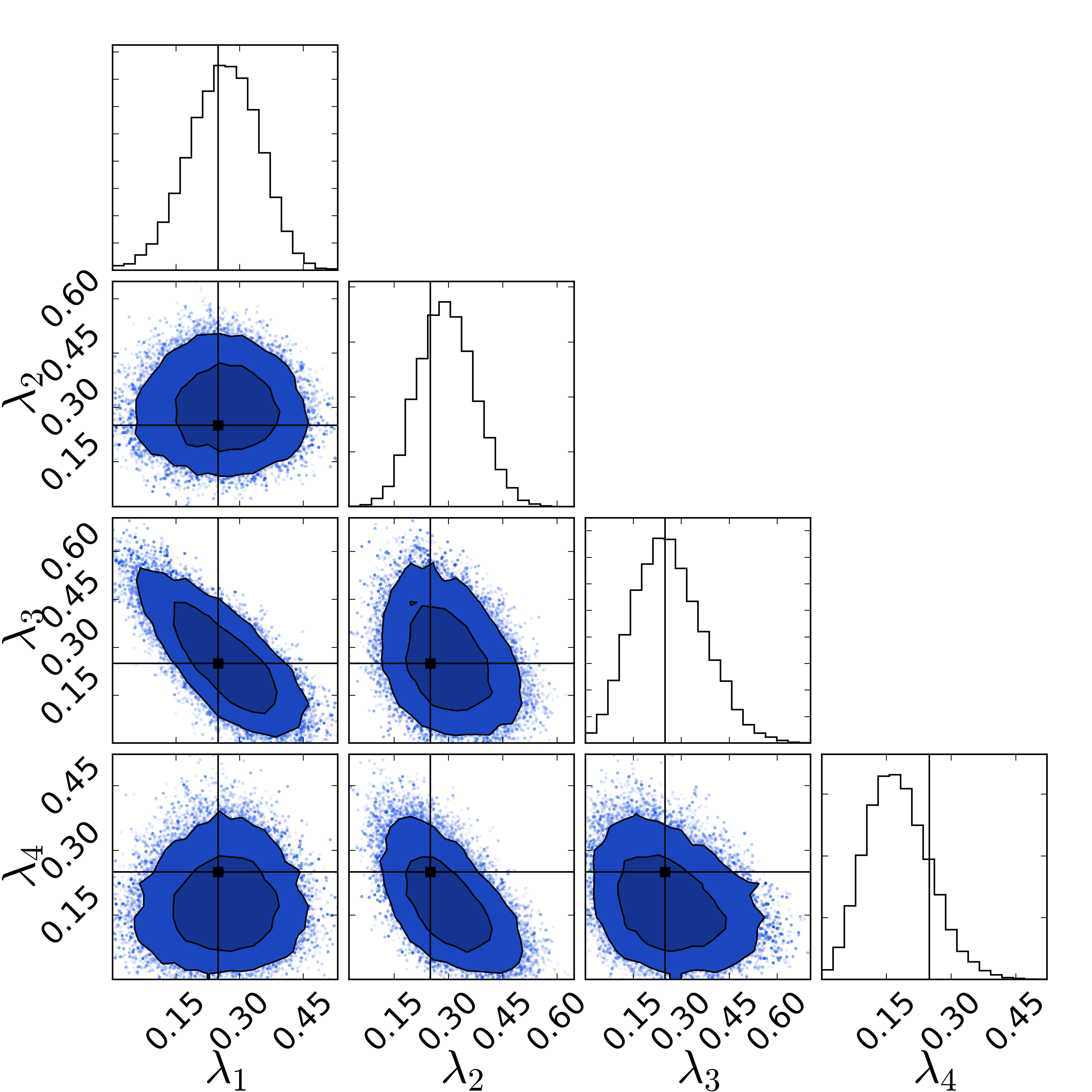} 
\caption{Marginalised 1D and 2D probability density functions for the $\lambda$ parameters describing the fractional contribution of each of the four subpopulations introduced in Section~\ref{sec:models}. The thin black lines indicate the true injection fraction from each model, which is $0.25$ for all models. The data used were the $80$ mock \ac{GW} events shown in Figure~\ref{fig:injections}, assumed to have perfect measurements of the spin--orbit misalignment angles $\cos{\theta_1}$ and $\cos{\theta_2}$. \change{Colours are the same as Figure~\ref{fig:dirichlet_prior}}} 

\label{fig:results_perfect_measurements_costheta1_costheta2}
\end{figure}

We sample Equation~\eqref{eq:posterior_lambda}, where our data consist of $80$ events with perfectly measured spin--orbit misalignments (as seen in Figure~\ref{fig:injections}). This number of detections could be available by the end of the third observing run under optimistic assumptions about detector sensitivity improvements  \citep{TheLIGOScientific:2016pea}. The results of this analysis are shown in Figure~\ref{fig:results_perfect_measurements_costheta1_costheta2}.

We find that after $80$ \ac{BBH} observations with perfect measurement accuracy, we would be able to confidently establish the presence of all four subpopulations. From this analysis, we can already understand some of the features of the posterior on the hyperparameters. For example, we see that there is a strong degeneracy between $\lambda_1$ and $\lambda_3$, since both of these models predict a large (nearly) aligned ($\theta_1 = \theta_2 = 0$) population. There is a similar degeneracy between $\lambda_2$ and $\lambda_4$. We can also see that the fraction of exactly aligned systems ($\lambda_1$) and the fraction of systems with isotropically distributed spin--orbit misalignments ($\lambda_2$) are not strongly correlated. Both fractions are measured with to be between $\sim 0.15$ and $\sim 0.45$ at the $90\%$ credible level with $80$ \ac{BBH} observations, corresponding to a fractional uncertainty of $\sim 50\%$. 

\subsection{Realistic measurement accuracy}

We know that in practice \ac{GW} detectors will not perfectly measure the spin--orbit misalignments of merging \acp{BBH} (see Figure~\ref{fig:typical_event_posterior_costheta1_costheta2} for a typical marginalised posterior). We now use the full set of $80$ \texttt{LALInference} posteriors, each containing $\sim 5000$ posterior samples as our input data, when sampling Equations~\eqref{eq:likelihood_sum_over_samples} and \eqref{eq:posterior_lambda}.

\begin{figure}
\centering
\includegraphics[width=0.45\textwidth]{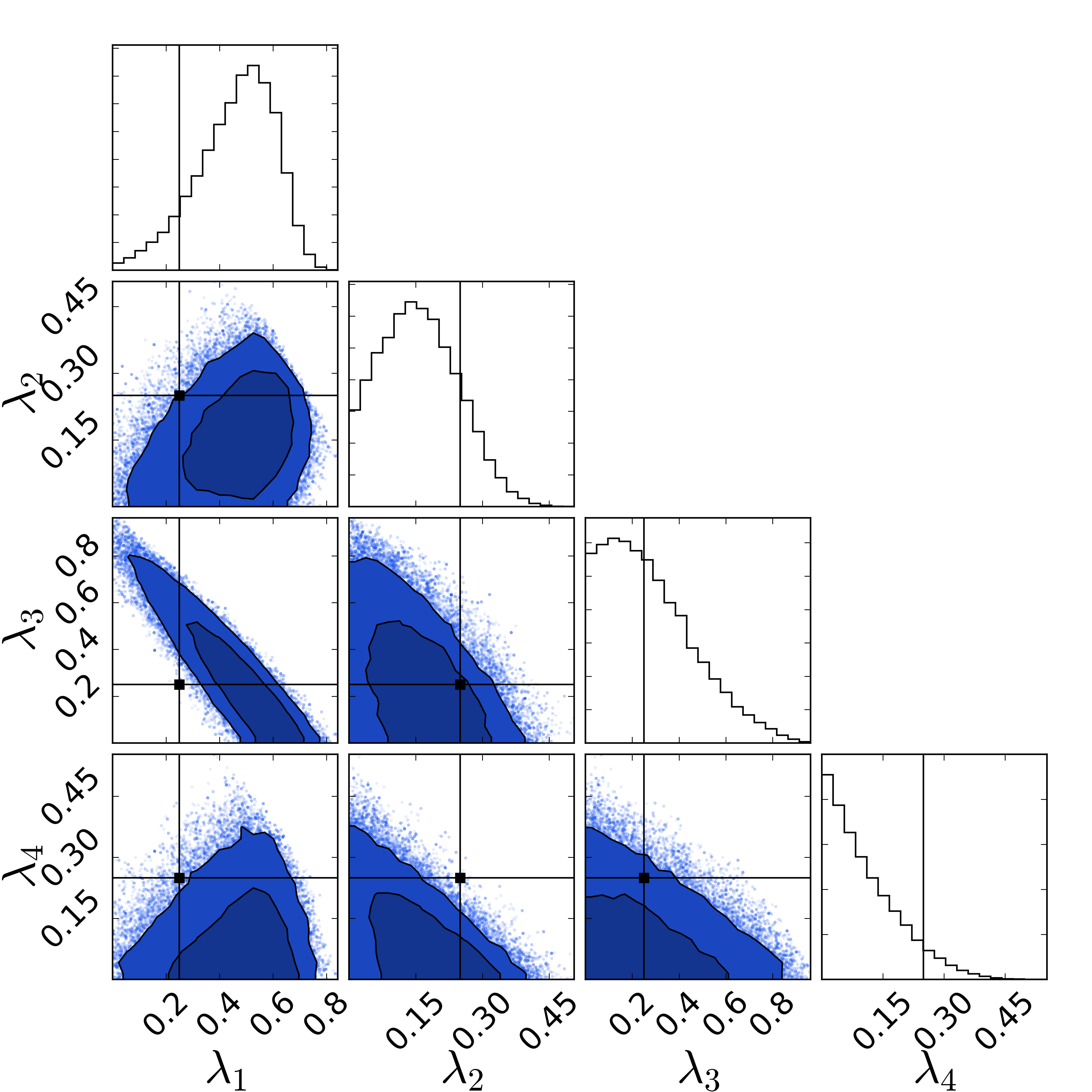} 
\caption{Marginalised 1D and 2D probability density functions for the $\lambda$ parameters describing the fractional contribution of each of the four subpopulations introduced in Section~\ref{sec:models}. The thin black lines indicate the true injection fraction from each model, which is $0.25$ for all models.  The data used were the full  \texttt{LALInference} posteriors of the $80$ mock \ac{GW} events shown in Figure~\ref{fig:injections}. \change{Colours are the same as Figure~\ref{fig:dirichlet_prior}}}
\label{fig:results_costheta1_costheta2}
\end{figure}

We show the results of this analysis in Figure~\ref{fig:results_costheta1_costheta2}. Many of the features seen in the posteriors on the hyperparameters are the same as those seen in Figure~\ref{fig:results_perfect_measurements_costheta1_costheta2}, such as the strong anti-correlation between $\lambda_1$ and $\lambda_3$. We see that the posterior is not perfectly centred on the true $\lambda$ values, though the true values do have posterior support. While the hierarchical modelling unambiguously points to the presence of multiple subpopulations, with no single subpopulation able to explain the full set of observations, the data no longer require all four subpopulations to be present.

We have checked that the structure of this posterior is typical given the limited number of observations and the large measurement uncertainties. In the next section we show that our posteriors converge to the true values in the limit of a large number of detections. 

\subsection{Dependence on number of observations}
\label{subsec:funcobs}

The \texttt{LALInference} \ac{PE} pipeline used to compute the posterior distributions for our 80 injections in Section~\ref{sec:pe} is computationally expensive. However, we would like to generate a larger catalogue of mock observations. First, this allows us to check that our analysis is self consistent by running many tests, such as confirming that the true result lies within the $P\%$ credible interval in $P\%$ of trials. Second, it allows us to predict how the accuracy of the inferred fractions of the subpopulations evolves as a function of the number of \ac{GW} observations.  

We develop approximations to these posteriors, similar to \citet{2017MNRAS.465.3254M}, based on the $80$ posterior distributions generated in Section~\ref{sec:pe}. The best measured spin parameter is a combination of the two component spins called the effective inspiral spin $\chi_\mathrm{eff} \in [-1, 1]$ \citep{2011PhRvL.106x1101A,TheLIGOScientific:2016wfe,Vitale:2015tea}:
\begin{equation}
\chi_\mathrm{eff} = \frac{\chi_{1} \cos{\theta_1} + q \chi_{2} \cos{\theta_2}}{(1 + q)} \, .
\label{eq:chi_eff}
\end{equation}
Having information about a single spin parameter makes it challenging to extract information about the spin distribution, but not impossible; for example, GW151226's positive $\chi_\mathrm{eff}$ means that at least one spin must have non-zero magnitude and $\theta_i < 90\degree$ \citep{Abbott:2016nmj}.

To compute the approximate posteriors, we represent each observation with true parameter values  $\chi_\mathrm{eff}^\mathrm{true}$ and $\cos{\theta_1}^\mathrm{true}$ by data which are maximum-likelihood estimates in a random noise realization:
\begin{align}
\chi_\mathrm{eff}^\mathrm{data} {} & {} \sim N\left(\chi_\mathrm{eff}^\mathrm{true}, \sigma^2_{\chi_\mathrm{eff}}(\chi_\mathrm{eff}^\mathrm{true})\right), \\
\cos{\theta_1}^\mathrm{data} {} & {} \sim N\left(\cos{\theta_1}^\mathrm{true}, \sigma^2_{\cos{\theta_1}}(\cos{\theta_1}^\mathrm{true})\right) \, ,
\end{align} 
where $N(\mu, \sigma^2)$ indicates a normal distribution. Posterior samples are then drawn using the same likelihood functions, centred on the maximum-likelihood data value,
\begin{align}
\chi_\mathrm{eff}^\mathrm{sample} {} & {} \sim N\left(\chi_\mathrm{eff}^\mathrm{data}, \sigma^2_{\chi_\mathrm{eff}}(\chi_\mathrm{eff}^\mathrm{sample})\right) ,\\
\cos{\theta_1}^\mathrm{sample} {} & {} \sim N\left(\cos{\theta_1}^\mathrm{data}, \sigma^2_{\cos{\theta_1} }(\cos{\theta_1}^\mathrm{sample})\right) \, .
\end{align} 
Here 
\begin{align}
\sigma_{\cos{\theta_1}} {} & {} = \frac{12}{\rho} (A \cos{\theta_1} + B)  ,\\ 
\sigma_{\chi_\mathrm{eff}} {} & {} = \frac{12}{\rho} C ,
\end{align}
with $A = -0.2$, $B = 0.3$ and $C = 0.2$ based on a fit to our $80$ posteriors; the measurement uncertainty scales inversely with the signal-to-noise ratio $\rho$ \citep{Cutler:1994ys}. 
In all cases, we only consider $\cos{\theta_1}$, $\cos{\theta_2}$ and $\chi_\mathrm{eff}$ in the permitted range of $[-1,1]$. 

\change{These posterior models are designed to mimic the typical properties of the posterior on the $\theta_1$ and $\theta_2$.  We have chosen a particular strategy that allows us to make faithful models; starting with models for posteriors on  $\chi_\mathrm{eff}$ and $\theta_1$ before converting them into posteriors on $\theta_1$ and $\theta_2$ is desirable because $\chi_\mathrm{eff}$ and $\theta_1$ are only weakly correlated, making it possible to write down independent expressions for the individual likelihoods, while $\theta_1$ and $\theta_2$ are strongly correlated, so would require a more complex joint likelihood model. The posteriors already implicitly account for marginalization over correlated parameters such as spin magnitudes and the mass ratio, which will not be known in practice. These mock posteriors will not correctly reproduce actual posteriors for specific events, since they do not include the noise realization, except probabilistically; they only reproduce the average properties of the posterior distributions.}

\begin{figure}
\centering
\includegraphics[width=0.45\textwidth]{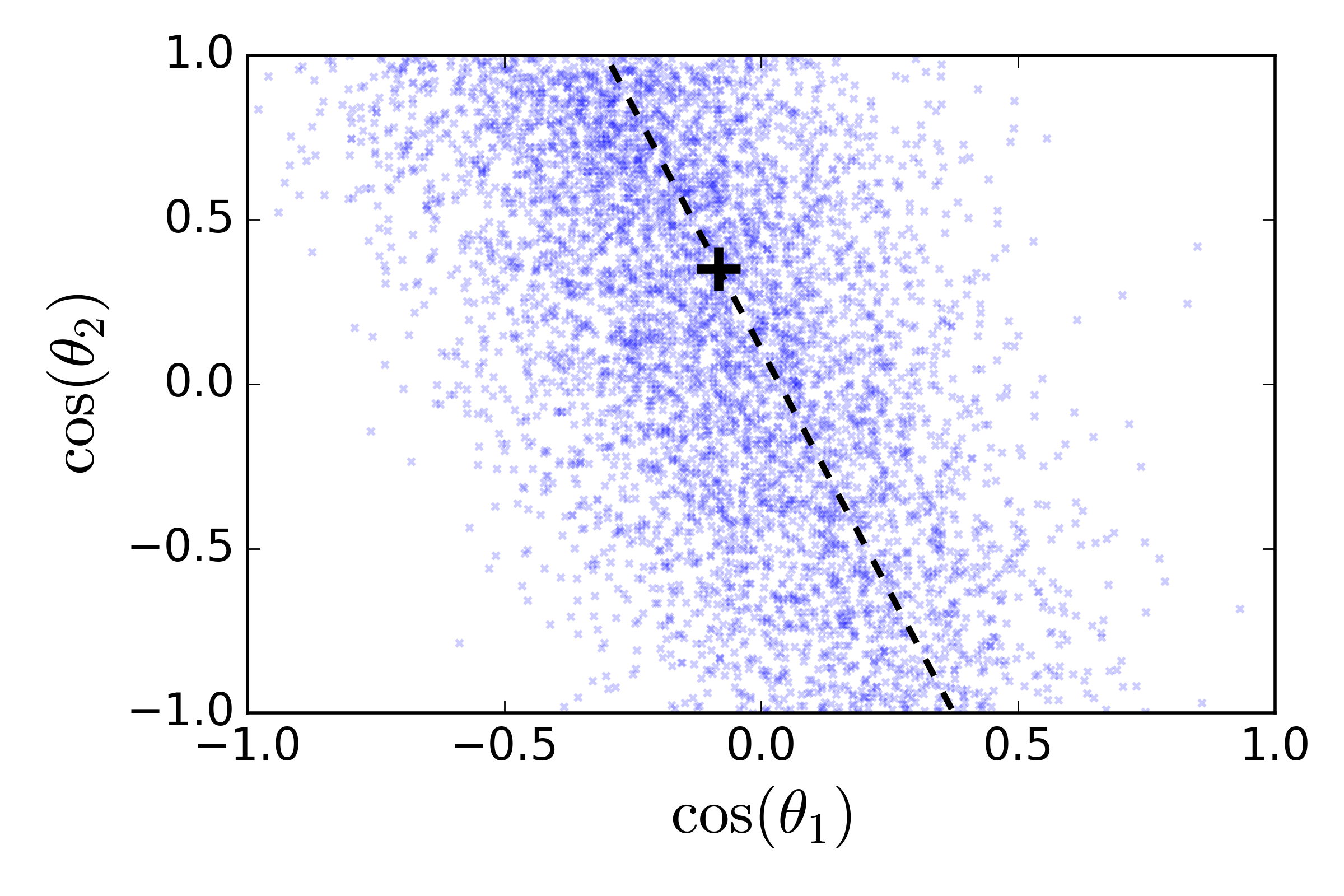} 
\caption{Marginalised posterior samples for the same event as shown in Figure~\ref{fig:typical_event_posterior_costheta1_costheta2}, with the same notation. This posterior distribution was approximated using the model described in section~\ref{subsec:funcobs}.}
\label{fig:typical_event_fake_posterior_costheta1_costheta2}
\end{figure}

We draw $\nu = 5000$ posterior samples of $\chi_\mathrm{eff}$ and $\cos{\theta_1}$ independently, and calculate the values of $\cos{\theta_2}$ using Equation~\eqref{eq:chi_eff}, fixing the mass ratio $q$ and spin magnitudes $\chi_i$ to their true values. This builds the correct degeneracies between $\cos{\theta_1}$ and $\cos{\theta_2}$ into the mock posteriors. We show an example of a posterior distribution generated this way in Figure~\ref{fig:typical_event_fake_posterior_costheta1_costheta2}.

\begin{figure}
\centering
\includegraphics[width=0.45\textwidth]{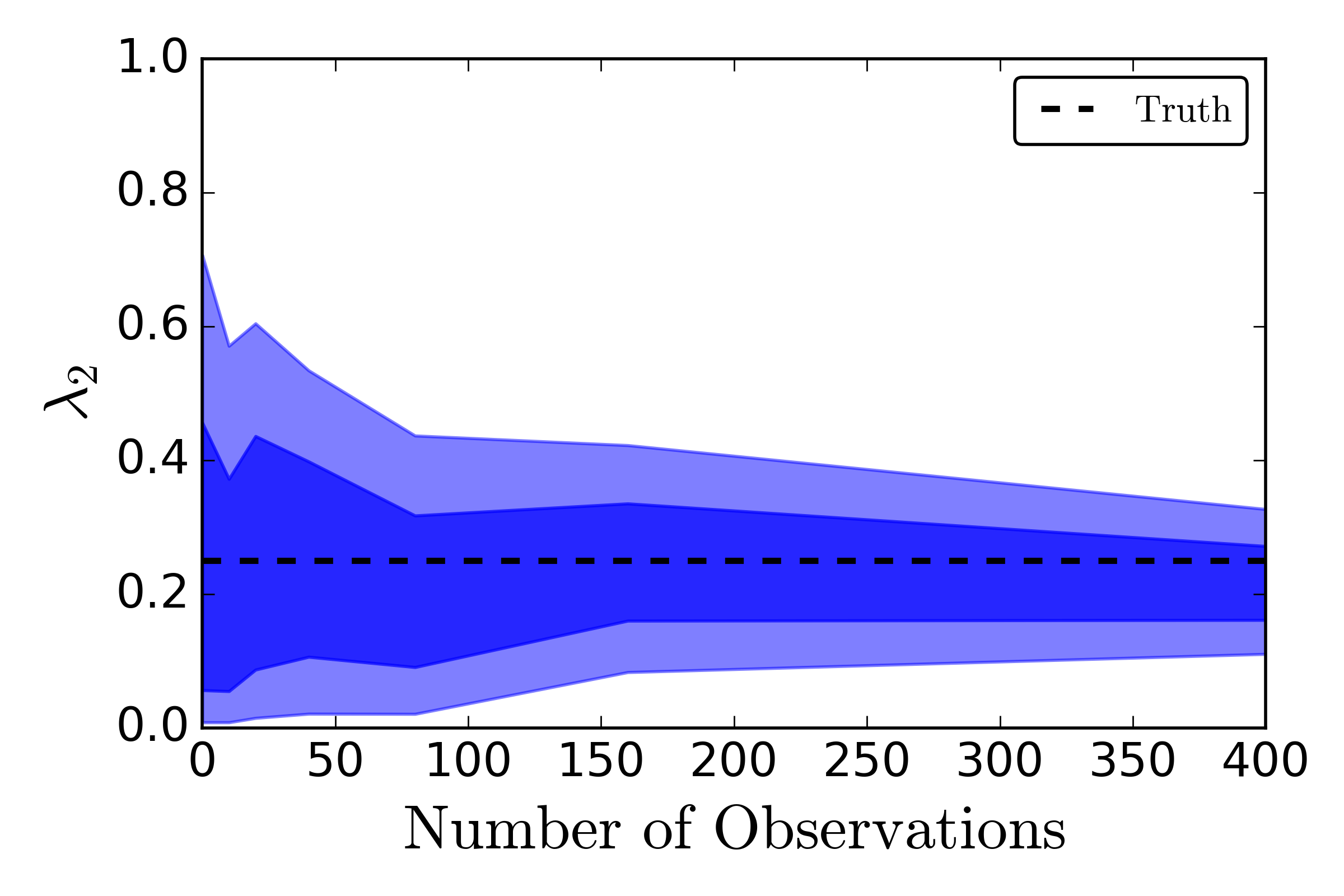}
\caption{Marginalised posterior on $\lambda_2$, the fraction of \acp{BBH} from the subpopulation with isotropic spin distribution (representing dynamical formation), as a function of the number of \ac{GW} observations. The posterior converges to the injected value of $\lambda_2 = 0.25$ (dashed black horizontal line) after $\sim 100$ observations. The coloured bands show the $68\%$ (darkest) and $95\%$ (lightest) credible intervals.}
\label{fig:function_of_observations_lambda_2}
\end{figure}
\begin{figure}
\centering
\includegraphics[width=0.45\textwidth]{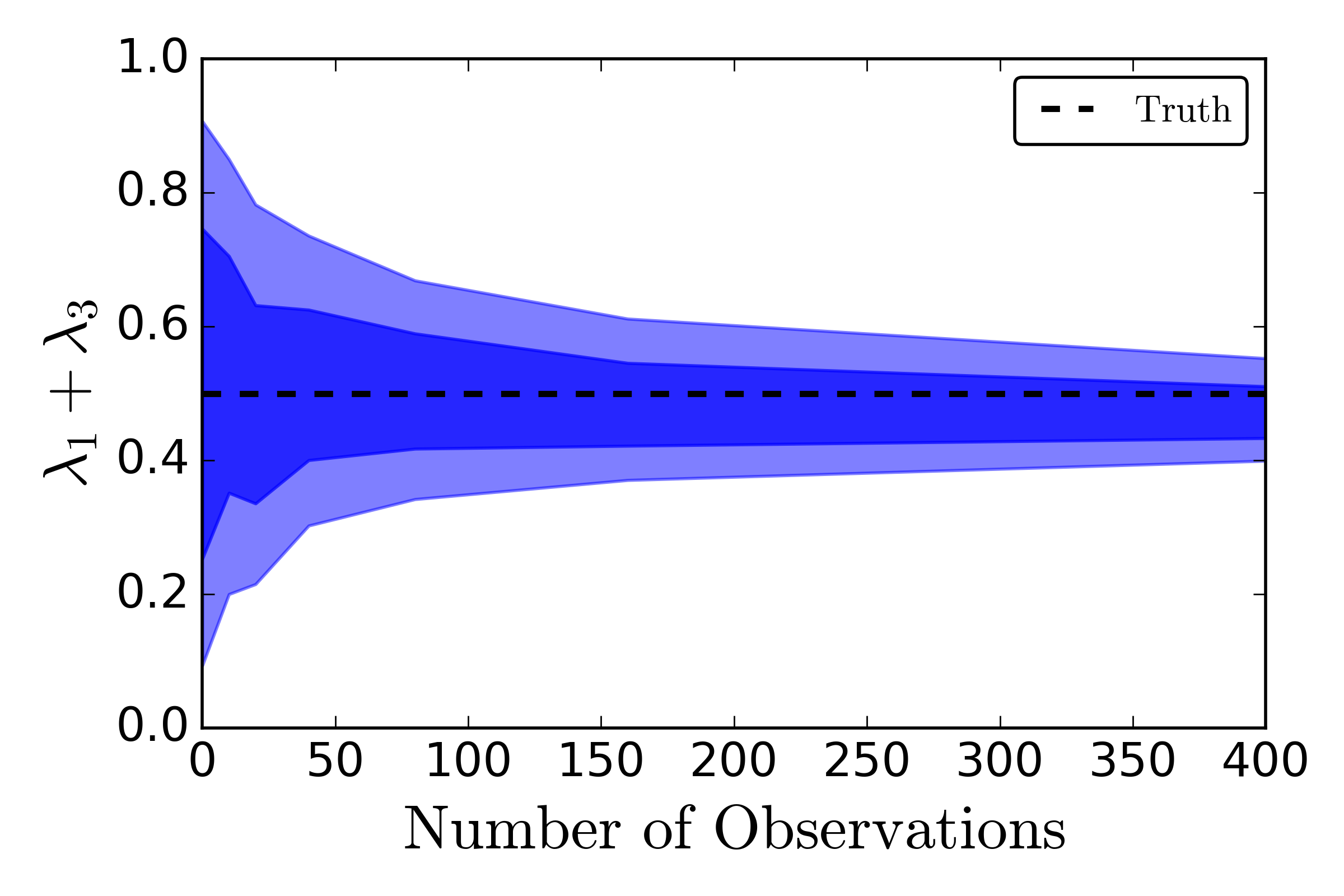}
\caption{Marginalised posterior on $\lambda_1 + \lambda_3$, the combined fraction of \acp{BBH} formed through subpopulations 1 and 3, as a function of the number of \ac{GW} observations. These subpopulations correspond to spins preferentially aligned with the orbital angular momentum. The posterior converges to the injected value of $\lambda_1 + \lambda_3= 0.5$ (dashed black horizontal line) after $\sim 20$ observations. The coloured bands show the $68\%$ (darkest) and $95\%$ (lightest) credible intervals.}
\label{fig:function_of_observations_lambda1_plus_lambda3}
\end{figure}

Using this method, we generate spin--orbit misalignment posteriors for $400$ \acp{BBH} drawn in equal fractions ($\lambda_i = 0.25$) from the four subpopulation models introduced in Section~\ref{sec:models}. Using the method introduced in Section~\ref{sec:hierarchical}, we calculate the posteriors on the $\lambda$ parameters after $0$ (prior), $10$, $20$, $40$, $80$, $160$ and $400$ observations, similar to \citet{2017MNRAS.465.3254M}. In Figure~\ref{fig:function_of_observations_lambda_2} we show the $68\%$ and $95\%$ credible intervals for the fraction $\lambda_2$ of observed \acp{BBH} coming from an isotropic distribution (subpopulation 2) corresponding to dynamically formed binaries. Given our models and incorporating realistic measurement uncertainties, we find that this fraction can be measured with a $\sim 40\%$ fractional uncertainty after $100$ observations. Since subpopulation 1 and 3 are somewhat degenerate in our model, we find that the combined fraction $\lambda_1 + \lambda_3$ is a well measured parameter (as shown in Figure~\ref{fig:function_of_observations_lambda1_plus_lambda3}), whilst the individual components are measured less well. For $N \gtrsim 100$ observations the uncertainties in the $\lambda_\ell$ scale as the inverse square root of the number of observations; e.g., $\sigma_{\lambda_1 + \lambda_3} \approx 0.8\, N^{-1/2}$. 

Although $\gtrsim 100$ observations are required to accurately measure the contribution of each of the four subpopulations, it is possible to test for more extreme models with fewer observations.  For example, $\sim 20$ 
observations are sufficient to demonstrate the presence of an isotropic subpopulation at the 95\% credible level.  

Even fewer observations are needed to confidently rule out the hypothesis that all observations come from the exactly aligned or isotropic subpopulations.  We draw observations from the isotropic subpopulation and calculate the ratio of the evidence (Bayes factor) $Z_\mathrm{aligned}$ for the model under which all \ac{BBH} spins are exactly aligned  ($\lambda_1 = 1$) to the evidence $Z_\mathrm{isotropic}$ for the model under which all \ac{BBH} spins are isotropically distributed ($\lambda_2 = 1$):
\begin{equation}
\frac{Z_\mathrm{aligned}}{Z_\mathrm{isotropic}} = \frac{p \left( \left\{ d_\alpha\right\}_{\alpha = 1}^{N} \middle| \lambda_1 = 1 \right)}{p \left( \left\{ d_\alpha\right\}_{\alpha = 1}^{N} \middle| \lambda_2 = 1 \right)}.
\end{equation}

\change{Inference using small numbers of observations is sensitive to the exact choice of these observations (both true parameters and measurement errors), which we randomly draw from the relevant distributions.}  We \change{therefore} repeat this test $100$ times to account for the random nature of the mock catalogue. In all cases, we find that with only $5$ observations of \acp{BBH} with component spin magnitudes $\chi = 0.7$, 
the exactly aligned model $\lambda_1 = 1$ can be ruled out at more than $5~\sigma$ confidence. Similarly, when drawing from the exactly aligned model, we find that the hypothesis that all events come from an isotropic population $\lambda_2 = 1$ can be ruled out at more than $5~\sigma$ confidence in all tests with $5$ observations.

\section{Discussion and conclusions}
\label{sec:conclusions}

With the first direct observations of \acp{GW} from merging \acp{BBH}, the era of \ac{GW} astronomy has begun. \ac{GW} observations provide a new and unique insight into the properties of \acp{BBH} and their progenitors. For individual systems, we can infer the masses and spins of the component \acp{BH}; combining these measurements we can learn about the population, and place constraints on the formation mechanisms for these systems, whether as the end point of isolated binary evolution or as the results of dynamical interactions.

In this work, we investigated how measurements of \ac{BBH} spin--orbit misalignments could inform our understanding of the \ac{BBH} population.  We chose the properties of our sources to match those we hope to observe with \ac{aLIGO} and \ac{AdV} (at design sensitivity), using four different astrophysically motivated subpopulations for the distribution of spin--orbit misalignment angles, each reflecting a different formation scenario.  We performed a Bayesian analysis of \ac{GW} signals (using full inspiral--merger--ringdown waveforms) for a population of \acp{BBH}.  We assumed a mixture model for the overall population of \ac{BBH} spin--orbit misalignments and combined the full \ac{PE} results from our \ac{GW} analysis in a hierarchical framework to infer the fraction of the population coming from each subpopulation. A similar analysis could be performed following the detection of real signals.  

\change{Adopting a population with spins of $\chi = 0.7$,} we demonstrate that the fraction of \acp{BBH} with spins preferentially aligned with the orbital angular momentum ($\lambda_1 + \lambda_3$) is well measured and can be measured with an uncertainty of $\sim 10\%$ with $100$ observations, scaling as the inverse square root of the number of observations.  We also show that after $100$ observations, we can measure the fraction $\lambda_2$ of the subpopulation with isotropic spins (assumed to correspond to dynamical formation) with a fractional uncertainty of $\sim 40\%$. 
Extreme hypotheses can be tested and ruled out with even fewer observations.  For example, with just $5$ observations we can rule out the hypothesis that all \acp{BBH} have their spins exactly aligned with high confidence ($>5~\sigma$) if the true population has isotropically distributed spins, and vice versa. This number of observations may be reached by the end of the second \ac{aLIGO} observing run.



One limitation of the current approach is the assumption that the subpopulation distributions are known perfectly. This will not be the case in practice, but the simplified models considered here are still relevant as parametrizable proxies for astrophysical scenarios. Hierarchical modelling with strong population assumptions could lead to systematic biases in the interpretation of the observations if those assumptions are not representative of the true populations; this can be mitigated by coupling such analysis with weakly modelled approaches, such as observation-based clustering \citep{2015MNRAS.450L..85M,2017MNRAS.465.3254M}.

In this work we have not taken into account observational selection effects, \change{particularly the differences in the detectabilty of different subpopulations because their different spin parameter distributions impact the \ac{SNR} \citep{2009PhRvD..80l4026R}}. These must be incorporated in the analysis \change{to correctly infer the intrinsic subpopulation fractions \citep{2013arXiv1302.5341F,Mandel:2016select}, but the impact is not expected to be significant in this case}. Care must also be taken to avoid biases when performing an hierarchical analysis with real observations, since the observations will not be drawn from the same distribution as the priors used for the analysis of individual events. Our framework accounts for the differences in the priors on the parameters of interest (spin--orbit misalignment angles) between the original \ac{PE} and model predictions, but not for any discrepancy in the priors of the parameters we marginalize over (e.g., masses); this is likely a second-order effect.

Neither theoretical models nor observations can currently place tight constraints on the spin magnitudes of \acp{BH}. We therefore chose to give all \acp{BH} a spin magnitude of $0.7$ in this study. This choice is clearly ad hoc; we expect a distribution of \ac{BH} spin magnitudes in nature. Since \ac{GW} events with \acp{BH} with low spin will not constrain the spin--orbit misalignment angles well, a distribution of spin magnitudes containing lower \ac{BH} spins will act to increase the requirements for the numbers of observations quoted here. \change{Even with significantly smaller \ac{BH} spin magnitudes, a few to a few tens of observations can distinguish between sufficiently different subpopulation models for spin--orbit misalignment (perfectly aligned versus isotropic) as shown in \citet{2017arXiv170601385F}.}

Here we have assumed that \acp{BH} receive large natal kicks, comparable to neutron stars, leading to relatively large spin--orbit misalignments even for isolated binary evolution. We further assume that the effect of the kick is simply to tilt the orbital plane, but not the \ac{BH} spin. There is, however, evidence from the Galactic double pulsar PSR J0737$-$3039 that the second born pulsar received a spin tilt at birth \citep{Farr:2011gs}. \change{Furthermore, alignment through mass transfer prior to \ac{BH} formation may be imperfect. Our models of \acp{BBH} that are preferentially but imperfectly aligned because of high natal kicks can be viewed as proxies for misalignment through a combination of these effects.}  

Optimal hierarchical modelling should fold in all available information, including component masses \change{\citep[cf.][]{2016ApJ...832L...2R}} and spin magnitudes \change{\citep[cf.][]{2017arXiv170306223G,2017arXiv170306869F}}, into a single analysis.  Complementary electromagnetic observations of high-mass X-ray binaries, Galactic radio pulsars, short gamma ray bursts, supernovae and luminous red novae will contribute to a concordance model of massive binary formation and evolution.

\section*{Acknowledgements}


This work was supported in part by the Science and Technology Facilities Council. IM acknowledges support from the Leverhulme Trust. 
We are grateful for computational resources provided by Cardiff University, and funded by an STFC grant supporting UK Involvement in the Operation of Advanced
LIGO. SS and IM acknowledge partial support by the National Science Foundation under Grant No.~NSF PHY11-25915.  We are grateful to colleagues from the Institute of Gravitational-wave Astronomy at the University of Birmingham, as well as Ben Farr, M.~Coleman Miller, Chris Pankow and Salvatore Vitale for fruitful discussions. We thank Davide Gerosa and the anonymous referee for comments on the paper. Corner plots were made using \texttt{triangle.py} available from \href{https://github.com/dfm/triangle.py}{https://github.com/dfm/triangle.py}. This is LIGO Document P1700007.
 



\bibliographystyle{mnras}
\bibliography{myBibliography}







\bsp	
\label{lastpage}
\end{document}